\documentclass[aps,twocolumn,showpacs,superscriptaddress,citeautoscript,reprint,
prb,floatfix]{revtex4-1}

\usepackage{bm}
\usepackage{graphicx}
\usepackage{amsmath}
\usepackage{amsfonts}
\usepackage{color}
\usepackage[colorinlistoftodos]{todonotes}
\usepackage{hyperref}
\usepackage{subfigure}

\hypersetup{
  pdftitle={Non-equilibrium transport through a disordered molecular nanowire},
  pdfauthor={P. Thiessen, F. Dominguez-Adame, R.A. Roemer, E. Diaz},
  pdfsubject={Transport through QDs},
  pdfstartview=FitH,
  breaklinks=true,
  bookmarksopen=true,
  bookmarksnumbered=true,
  unicode=true
}

\usepackage{color}
\definecolor{Blue}{rgb}{0.3,0.3,0.9}
\definecolor{Red}{rgb}{0.9,0.3,0.3}
\definecolor{Green}{rgb}{0.3,0.6,0.3}
\newcommand{\revision}[1]{\textcolor{Red}{{#1}}}

\newcommand{\rmi}{\mathrm{i}}

\newcommand{\Tr}{\mathrm{Tr}\,}

\begin{document}

\title{Non-equilibrium transport through a disordered molecular nanowire}

\author{P.\ Thiessen}

\affiliation{GISC, Departamento de F\'{\i}sica de Materiales, Universidad 
Complutense, E-28040 Madrid, Spain}

\author{E.\ D\'{\i}az}\email{elenadg@fis.ucm.es}

\affiliation{GISC, Departamento de F\'{\i}sica de Materiales, Universidad 
Complutense, E-28040 Madrid, Spain}

\author{R.A.\ R\"{o}mer}

\affiliation{Department of Physics and Centre for Scientific Computing, 
University of Warwick, Coventry, CV4 7AL, United Kingdom}

\author{F.\ Dom\'{\i}nguez-Adame}

\affiliation{GISC, Departamento de F\'{\i}sica de Materiales, Universidad 
Complutense, E-28040 Madrid, Spain}

\affiliation{Department of Physics and Centre for Scientific Computing, 
University of Warwick, Coventry, CV4 7AL, United Kingdom}


\begin{abstract}

We investigate the non-equilibrium transport properties of a disordered 
molecular nanowire. The nanowire is regarded as a quasi-one-dimensional organic 
crystal composed of self-assembled molecules. One orbital and a single 
\emph{random} energy are assigned to each molecule while the intermolecular 
coupling does not fluctuate. Consequently, electronic states are expected to be 
spatially localized. We consider the regime of strong localization, namely, the 
localization length is smaller than the length of the molecular wire. 
Electron-vibron interaction, taking place at each single molecule, is also 
considered. We investigate the interplay between static disorder and electron-vibron 
interaction in response to either an applied electric bias or a temperature 
gradient. To this end, we calculate the electric and heat currents when the 
nanowire is connected to leads, using the Keldysh non-equilibrium Green's 
function formalism. At intermediate temperature, scattering by disorder 
dominates both charge and heat transport. We find that the electron-vibron 
interaction enhances the effect of the disorder on the transport properties due 
to the decrease of the coherent electron tunneling among molecules.

\end{abstract}

\maketitle

\section{Introduction}

Anderson localization of the electronic wave function in a random medium is a major paradigm of quantum 
coherence in condensed matter physics: non-interacting electrons in three dimensions are spatially localized for 
sufficiently large disorder\cite{Anderson58} and in one-dimension all the states of random systems become exponentially localized 
for any amount of disorder due to coherent backscattering~\cite{Mott61}. With a few 
 exceptions~\cite{Kohmoto1986,DeMoura1998,Bellani99,Rodriguez00,Kuhl00,Queiroz02,Rodriguez12}, a 
single-parameter scaling theory\cite{Abrahams79} generally provides a very accurate picture of 
the electronic states in non-interacting disordered systems. 
In real solids, however, electrons interact with each other and with lattice vibrations, and these interactions may affect the transport properties of disordered systems. For instance, electron-phonon interaction can decrease the ability of electrons to form localized states and hence increase charge mobility~\cite{Kopidakis96,Bronold04}.


The advent of nanotechnology has renewed attention on Anderson 
localization because it is enhanced in low-dimensional system~\cite{Abrahams79}. 
Among the large variety of
materials with technological interest in this field, crystalline molecular systems 
are gaining relevance as active components in electronic nanodevices.\cite{Bakulin2015} 
Unfortunately,
the detailed mechanisms of charge transport in molecular systems driven out of 
equilibrium are still controversial, posing a complicated scenario for the 
theoretical description of experiments.\cite{Erpenbeck2015} 
For example, it has been argued that, depending on the various energy
scales involved (electron bandwidth, zero-point energy of molecular vibrations, 
thermal energy), electron-phonon coupling may not play a significant role on 
charge transport even at room temperature, as deduced from inelastic electron 
tunneling spectroscopy experiments~\cite{Wang04,Galperin04,Kubatkin09,Lykkebo13}. 
When the charge carriers interact with low-energy intermolecular modes, they
move in a slowly-changing potential landscape that gives rise to the so called
\emph{transient localization}~\cite{Fratini16}. On the other hand, intramolecular 
modes occur at high frequency due to the stretching of stiff covalent bonds. 
Coupling to those modes may strongly alter charge transport~\cite{Nan09}
and even lead to the self-trapping of charge carriers, provided that the
relaxation energy (the energy gained upon the deformation of lattice around
the carrier) largely exceeds the kinetic energy gained from the carrier 
tunneling to neighbor molecules~\cite{Fratini16}. 



In this work we aim at exploring the intermediate regime when the electron-vibron 
interaction is not strong and the effect on the transport properties of a long and
 pristine molecular nanowire~(MNW) are expected to be small. We are particularly interested 
in the \revision{interplay between static disorder and polaronic effects on the charge and heat transport properties in this regime}. 
The MNW will be regarded 
as a quasi-one-dimensional organic crystal of self-assembled 
molecules~\cite{Wang11}. Specifically, we consider a MNW with the two ends 
connected to ideal leads and assume that the electrons interact with a 
vibrational degree of freedom localized at each molecule. Disorder in the 
electronic environment of each molecule can originate from interactions with a 
random environment of solute molecules and ions surrounding the MNW. Unlike 
Ref.~\onlinecite{Amato90}, we propose a two-probe configuration to diminish the 
conductance fluctuations found in four-probe setups.  It should be stressed that 
we neglect the lateral motion of electrons in the MNW. Grange has recently 
established that the current-voltage characteristics shows a transition 
from wide to narrow wires, displaying additional peaks due to resonances with 
optical phonons~\cite{Grange14}. Since we only deal with quasi-one-dimensional 
MNWs, this transition is beyond the scope of our work. We will use the Keldysh 
non-equilibrium Green's function formalism~\cite{Haug07} to obtain the 
spectral function as well as the electric and heat currents through the MNW, driven 
out of equilibrium by either an applied electric bias or temperature gradient. 

One of our main findings is the strong effect of disorder on the electron transport 
properties of the MNW when the electron-vibron interaction is taken into account. 
The enhancement of the localization effects can be traced back to the so-called 
\emph{exponential suppression of tunneling}~\cite{Rudzinski08}. This amounts to reducing 
the coherent electron tunneling among neighbor molecules when the electron-vibron interaction 
is non-negligible. Consequently, the ratio between the magnitude of disorder, defined as 
the width of the distribution of site energies, and the electron hopping parameter among 
neighbor molecules becomes larger for higher temperatures. This can actually be viewed 
as an \emph{effective increase} of the disorder. Notice that, due to the zero-point motion, 
the hopping parameter is already significantly decreased at $T=0$, leading to a stronger 
effective disorder even in the absence of background temperature. 

\begin{figure}[tb]
\begin{center}
\includegraphics[width=0.9\linewidth]{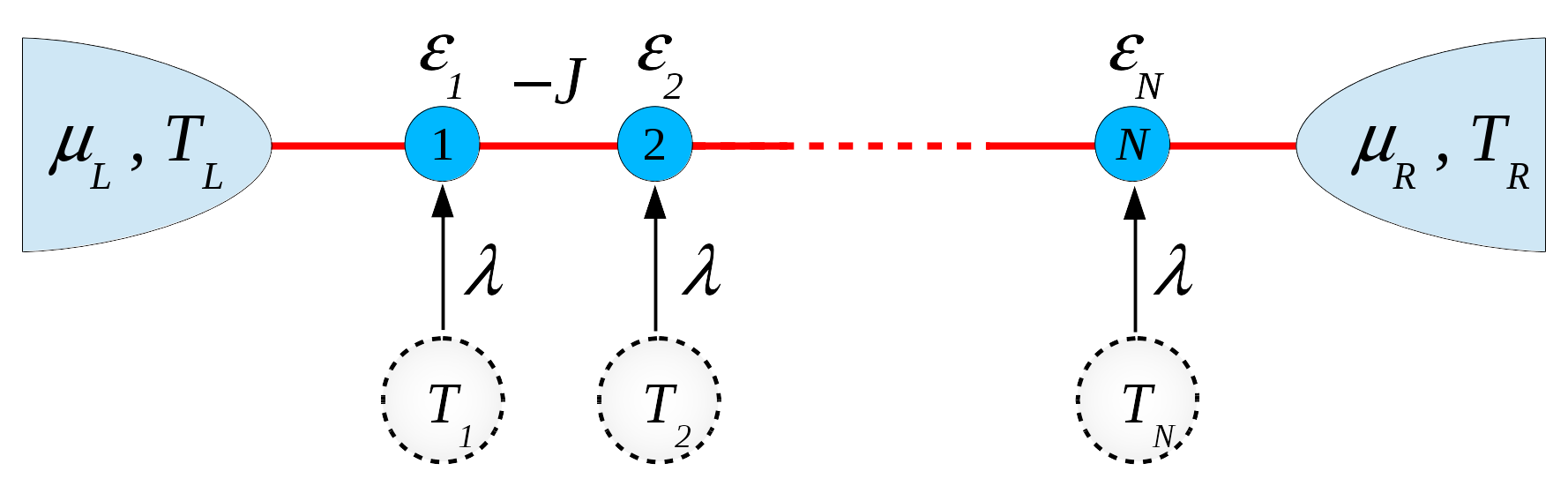}
\end{center}
\caption{(Color online) Schematic diagram of the MNW with intermolecular hopping 
parameter $J$. The MNW is also connected to left and right leads. We take into 
account an intramolecular electron-vibron interaction, $\lambda$ being the 
coupling constant and $T_i$ the temperature of the bath for each molecule. The $\varepsilon_i$ denote the local molecular energies and $\mu_{L,R}$ and $T_{L,R}$ are the chemical potentials and the temperatures in the left and right leads, respectively.}
\label{fig1}
\end{figure}
 
\section{Model and formalism}

We consider a MNW composed of $N$ self-assembled molecules and connected to 
left~($L$) and right~($R$) leads by tunneling couplings, as shown schematically in 
Fig.~\ref{fig1}. The chemical potentials of the leads under a bias voltage $V$ 
are given by $\mu_L=\mu + eV/2$ and $\mu_R=\mu - eV/2$, where $\mu$ is the 
equilibrium chemical potential and $e$ the electron charge. 
The lead temperatures are set as $T_L=T-\Delta T/2$ and $T_R=T+\Delta T/2$, where $T$ is a background temperature and $\Delta T$ is 
the temperature difference between the hot and the cold leads. 

Only one energy level in each molecule is assumed relevant and electron-electron 
interaction is neglected. On-site energies are subjected to disorder 
representing inhomogeneous broadening. Then, the energy level of the 
molecule~$i$ splits as $\varepsilon_i=\varepsilon + \Delta\varepsilon_i$, where 
$\Delta\varepsilon_i$ is a random uncorrelated variable whose distribution 
function is $P(\Delta\varepsilon_i)=1/W$ if $\vert\Delta\varepsilon_i\vert < 
W/2$ and zero otherwise. $W$ will be referred to as magnitude of disorder.
In addition, the electron interacts with a local vibration mode at each 
molecule which we assume of the same frequency $\omega_0$ for simplicity. 
Besides the different temperature of the leads, we introduce a temperature 
gradient in the system by setting a different temperature $T_i$ of each bath 
(see Fig.~\ref{fig1}). In this work we interpolate linearly $T_i$ between $T_L$ 
and $T_R$. 
 
\subsection{The coupled electron-vibron system}

The Hamiltonian describing the whole system splits into three contributions as 
$H=H_{0}+H_\mathrm{e-leads}+H_\mathrm{e-vib}$~\cite{Alvarez15}. The term $H_{0}$ describes the 
dynamics of the non-interacting system (we set $\hbar=1$)
\begin{align}
H_0&=\sum_{i=1}^{N} \varepsilon_{i}^{}c_{i}^{\dagger}c_{i}^{}
-J\sum_{i=1}^{N-1}
\left(c_{i}^{\dagger}c_{i+1}^{}+c_{i+1}^{\dagger}c_i^{}\right)
\nonumber\\
&+\omega_0 \sum_{i=1}^{N} a_{i}^{\dagger}a_{i}^{}+\sum_{\alpha k} 
\varepsilon_{\alpha k}^{}d_{\alpha k}^{\dagger}d_{\alpha k}^{}\ .
\label{eq:01}
\end{align}
Here $d_{\alpha k}^{\dagger}$ ($d_{\alpha k}^{}$) denotes the creation 
(annihilation) operator of a conduction electron in the lead $\alpha=L,R$ with 
 crystal momentum $k$ and energy $\varepsilon_{\alpha k}$. Similarly, 
$c_{i}^{\dagger}$ ($c_{i}^{}$) is the creation (annihilation) operator of an 
 electron in the molecule~$i$ with energy $\varepsilon_{i}$. $J$ indicates the bare
intermolecular hopping energy and is assumed constant and positive. It is worth 
noticing that $J$ depends on the particular distance and orientation between neighboring 
units and it could change due to molecular vibrations~\cite{Coropceanu07}. However this 
issue is beyond our current study and further works should consider it. Finally, the 
creation (annihilation) operator of a vibron in the molecule~$i$ with frequency 
$\omega_0$ is denoted by $a_{i}^{\dagger}$ ($a_{i}^{}$).

The MNW is tunnel-coupled to both leads, as shown schematically in 
Fig~\ref{fig1}. Therefore, the corresponding Hamiltonian reads
\begin{equation}
H_\mathrm{e-leads} = \sum_{\alpha ki} \left(V_{\alpha ki}^{}d_{\alpha 
k}^{\dagger}
c_i^{} + V_{\alpha ki}^{*}c_i^{\dagger}d_{\alpha k}^{}\right)\ .
\label{eq:02}
\end{equation}
\revision{Self-trapping has been commonly formulated within the framework of the small 
polaron theory based on a local Holstein-type coupling~\cite{Holstein59} between 
the carrier and the intramolecular modes. Quanta of the intramolecular vibrations 
are usually referred to as vibrons. The Holstein-type coupling between the electron and the 
vibrons~\cite{Holstein59} can be written as}
\begin{equation}
H_\mathrm{e-vib} = \lambda \sum_{i}\left(a_{i}^{\dagger}+a_{i}^{}\right)
c_i^{\dag}c_i^{}\ .
\label{eq:03}
\end{equation}
The electron-vibron coupling constant $\lambda$ is assumed uniform over the MNW.
We now apply the polaron (Lang-Firsov{\cite{Lang63}) nonperturbative canonical transformation
when the coupling to the leads is not strong ($|V_{\alpha k i}|< \lambda$). 
In such a case, it is reasonable to replace the displacement 
operator $X_i=\exp[-(\lambda/\omega_0)(a_i^{\dagger}-a_i^{})]$ that emerges after 
the transformation by its thermal expectation value evaluated in equilibrium 
$\langle X_i \rangle$ . Notice that such polaron transformation 
and the subsequent replacement turns the original many-body problem into an 
effective one-body problem (see Appendix~\ref{sec:appA} for details). The transformed 
Hamiltonian is approximately given by
\begin{align}
\widetilde{H}&\approx\sum_{i=1}^{N}\widetilde{\varepsilon}_i^{}
c_i^{\dagger}c_i^{} - \sum_{i=1}^{N-1} 
\left(\widetilde{J}_ic_{i}^{\dagger}c_{i+1}^{} + \textrm{H.c.}\right)
+\sum_{i=1}^{N} \omega_{0}^{} a_i^{\dagger}a_i^{} \nonumber \\
&+\sum_{\alpha ki} \left(\widetilde{V}_{\alpha k i}^{}
d_{\alpha k}^{\dagger}c_i^{} + \textrm{H.c.}\right) \ ,
\label{eq:05}
\end{align}
with $\widetilde{\varepsilon}_i=\varepsilon_i-\lambda^2 /\omega_0$ being
the renormalized energy level of the molecule~$i$, $\widetilde{J}_i=
J\exp\left[-\xi_i(T_i)/2-\xi_{i+1} (T_{i+1})/2\right]$ and $\widetilde{V}_{\alpha ki}^{}
=V_{\alpha ki}^{}\exp[-\xi_i(T_i)/2]$. Here H.c. stands for Hermitian conjugate. We have defined 
$\xi_i(T_i)=g\left(2n_{i}+1\right)$, $n_{i}=1/\left[\exp\left(\omega_0/k_{B}T_i\right)-1\right]$ 
and $g=\lambda^2/\omega_0^2$. Notice that 
the higher the temperature, the smaller the dressed couplings among the 
molecules $\widetilde{J}_i$ and between them and the leads $\widetilde{V}_{\alpha ki}$, as pictured in 
Fig.~\ref{fig:J_T}. 
This will be referred to as exponential suppression of 
tunneling~\cite{Rudzinski08}. It is worth mentioning that 
$\widetilde{J}_{i}/J<1$ at $T=0$ due to the zero-point motion.

\begin{figure}[tb]
\centering
\includegraphics[width=0.9\linewidth]{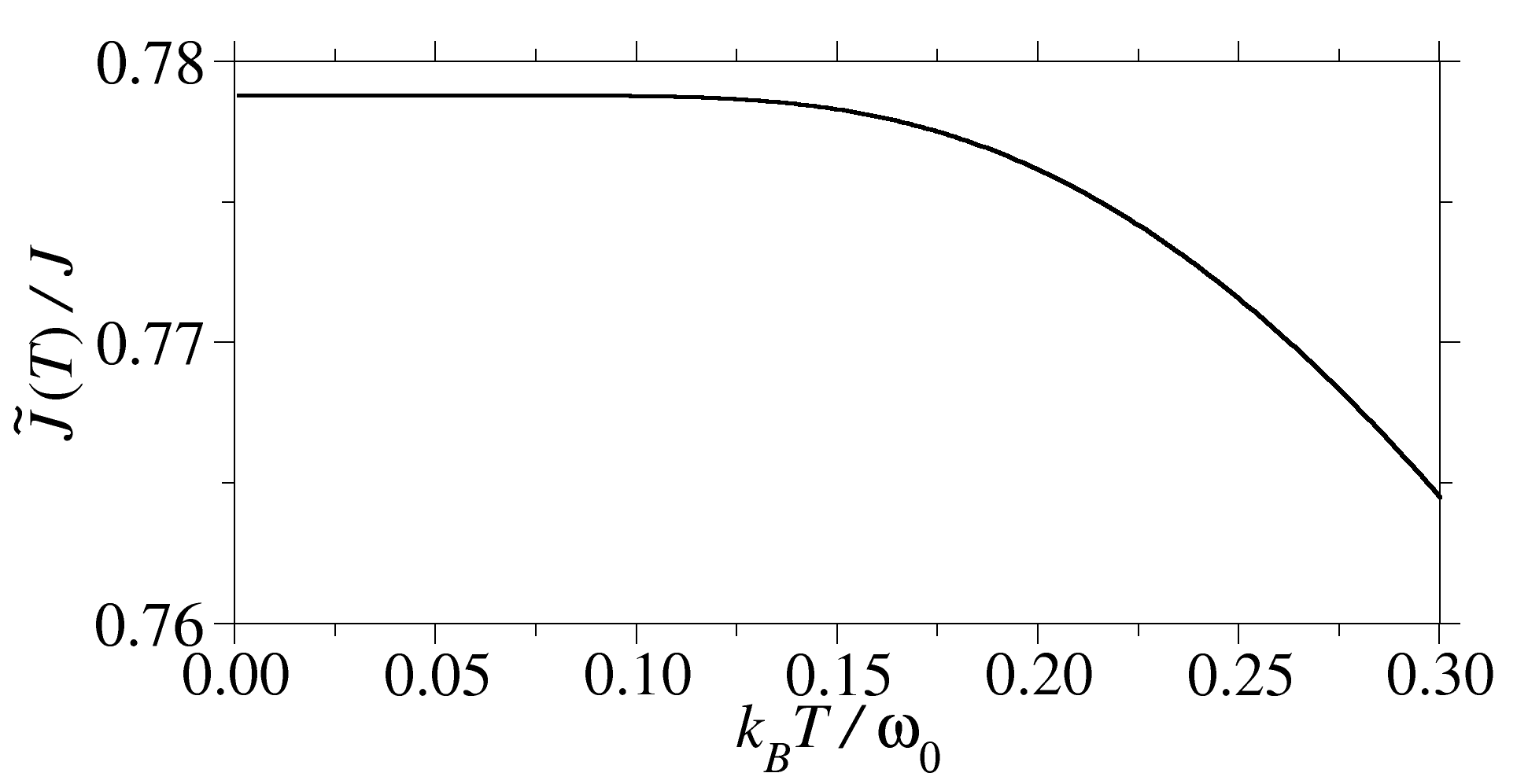}
\caption{Temperature dependence of the dressed hopping energy 
$\widetilde{J}$ among the molecules at an electron-vibron coupling strength 
of $\lambda=0.25$, in units of the bare hopping energy $J$. In the absence of 
temperature gradient the dressed hopping energy is independent of the site 
index. The observed monotonic decay continues for larger $T$ values.}
\label{fig:J_T}
\end{figure}

\subsection{Non-equilibrium transport properties}

Non-equilibrium transport properties of an interaction region coupled to two 
leads can be obtained with the help of the Keldysh non-equilibrium Green's 
function technique~\cite{Keldysh65}. This procedure is detailed in Appendix B. 

After transformation into Fourier space, the calculation of 
the greater and lesser Green's functions, $G^>(\omega)$ and $G^<(\omega)$, allows us to obtain the spectral matrix
\begin{equation}
\mathcal{A}(\omega)=\rmi \big[ G^>(\omega)-G^<(\omega) \big]\ ,
\label{eq:08}
\end{equation}
along with the spectral function 
$A(\omega)=\Tr\left[\mathcal{A}(\omega)\right]/N$. 
In addition, we can also calculate the 
symmetrized electric current~\cite{Haug07}
\begin{subequations}
\begin{eqnarray}
J_e&=&\frac{e}{2}\int \frac{d\omega}{2\pi} \, \Tr \Big[(\Gamma^L-\Gamma^R)\rmi 
G^{<}(\omega)
\nonumber\\
&+&\Big(f_L(\omega)\Gamma^L-f_R(\omega)\Gamma^R\Big)\mathcal{A}(\omega)\Big]\ .
\label{eq:18a}
\end{eqnarray}
Here 
$f_\alpha(\omega)=1/\left\{\exp\left[(\omega-\mu_\alpha)/k_{B}T_{\alpha}
\right] +1 \right\}$ is the Fermi-Dirac distribution function of the lead 
$\alpha$. The matrices which encode 
the coupling to the leads, $\Gamma^{\alpha}$, are taken symmetric with 
$\Gamma^{L}_{1,1}=\Gamma^{R}_{N,N}$ ($\Gamma^{\alpha}_{ij}=0$ otherwise).
Notice that we will neglect their $k$-dependence by relying on the wide-band limit approximation and take these matrices as energy-independent magnitudes. 

Unlike the electric current, the heat current is not necessarily conserved due 
to the coupling to the heat baths and Joule heating. We will mainly concentrate 
on the heat current from the left lead to the MNW 
\begin{eqnarray}
J_Q^L&=&\int \frac{d\omega}{2\pi}\, \omega \,\Tr \Big[\Gamma^L \rmi 
G^{<}(\omega)
+f_L(\omega)\Gamma^L\mathcal{A}(\omega)\Big]
\nonumber\\
&-& \mu_L\,\frac{J_e}{e} ,
\label{eq:18b}
\end{eqnarray}
\label{eq:18}
\end{subequations}
which is expected to differ from the heat current $J_{Q}^{R}$ from the right 
lead to the MNW. For instance, due to the symmetry of the system at $\Delta T=0$, the heat current $J_{Q}^{L}(-\Delta T)$ from the left lead to the system at $-\Delta T$ is equal to the energy flux $J_{Q}^{R}(\Delta 
T)$ from the right lead to the system at $\Delta T$. Following the reasoning 
from Ref.~\onlinecite{Ratner07} the sum $J_{\Delta E}(\Delta T)\equiv 
J_{Q}^{L}(\Delta T)+J_{Q}^{R}(\Delta T)=J_{Q}^{L}(\Delta T)+J_{Q}^{L}(-\Delta 
T)$ equals the rate of energy generated inside the MNW. 
In the following sections the electric and heat currents will 
be expressed in units of $e\omega_{0}/2\pi$ and $\omega_{0}^{2}/2\pi$, 
respectively. The superscript $L$ in Eq.~(\ref{eq:18b}) will be removed unless 
stated otherwise.

\section{Spectral function \label{sec:spectral}}

In this section we present and discuss the salient features of the spectral 
function $A(\omega)$. This quantity provides information of the energy spectrum of 
the elementary excitations in the system. For concreteness we restrict ourselves to 
the equilibrium regime by setting $\mu_L=\mu_R=0$ and $\Delta T=0$ throughout this 
section. To gain insight into the effect of the electron-vibron interaction we compare 
the numerically calculated spectral density in the non-interacting MNW ($\lambda=0$) 
with a wire where the electron-vibron coupling strength is finite 
($\lambda=0.5$). Energies are expressed in units of the vibron energy $\omega_0$ 
in what follows (recall that we set $\hbar=1$). As already mentioned, the 
coupling to the leads is taken as symmetric with 
$\Gamma^{L}_{1,1}=\Gamma^{R}_{N,N}=0.2$ ($\Gamma^{\alpha}_{ij}=0$ otherwise),
the bare intermolecular hopping energy is $J=0.1$ and $N=20$. 

Figure~\ref{fig:spectral}(a) shows the spectral density of a uniform MNW ($W=0$) 
with constant on-site energy $\varepsilon_{0}=0.25$ (blue solid line) and 
$\varepsilon_{0}=0$ (red solid line), which corresponds to renormalized energies 
$\tilde{\varepsilon}_{0}=0$ and $\tilde{\varepsilon}_{0}=-0.25$ when 
$\lambda=0.5$, respectively. The temperature of the system is 
$k_{B}T=0.1$. It is important to stress that we are assuming that the system is gated
so the energy level $\varepsilon_0$ can be set at the chemical potential of the contacts
at equilibrium $\mu$~\cite{Poot06}. In addition, the Huang-Rhys factor $g=\lambda^2/\omega_0^2$
turns out to be $g=0.25$, which falls in the typical range of parameters of 
organic semiconductors~\cite{Ueno08}. Results are compared to the non-interacting case 
($\lambda=0$, black dashed line) when $\varepsilon_{0}=0$, whose spectral 
density displays the expected $U$-shaped profile of width $4\widetilde{J}$ 
corresponding to a one-dimensional lattice with dressed intermolecular hopping 
energy $\widetilde{J}$. When the electron-vibron interaction is turned on (red 
solid line), the zero-vibron band of the spectral density is red-shifted with 
regard to the non-interacting case, according to the renormalization of the 
on-site energy $\tilde{\varepsilon}_{0}=\varepsilon_{0}-\lambda^{2}=-0.25$ 
associated to the deformation of the lattice around the tunneling 
electron~\cite{Wingreen88}. Furthermore, the finite electron-vibron interaction 
leads to the formation of side bands centered at energies 
$\tilde{\varepsilon}_{0}-1$ and $\tilde{\varepsilon}_{0}+1$, which correspond to 
emission or absorption of vibrational energy, respectively. As the thermal 
energy is small compared to the vibrational energy ($k_{B}T\ll 
\omega_{0}$), the latter contribution is weak, which leads to an asymmetric 
spectral density $A(\omega)$ and therefore to a broken particle-hole symmetry in 
the case $\widetilde{\varepsilon}_{0}\neq 0$. For the case of a renormalized 
energy of $\widetilde{\varepsilon}_{0}=0$ (blue solid line), the particle-hole 
symmetry is established and the side bands are symmetric. It has to be noted 
that the sum rule$\int\text{d}\omega\mathcal{A}(\omega)=2\pi$ still holds, so 
that the additional contribution due to the side channels are compensated by 
decreasing the height of the zero-vibron band with respect to the 
non-interacting case, as seen when comparing the blue solid line with the black 
dashed line~\cite{Chen05}.
%
\begin{figure}[tb]
\centerline{\includegraphics[width=\columnwidth]{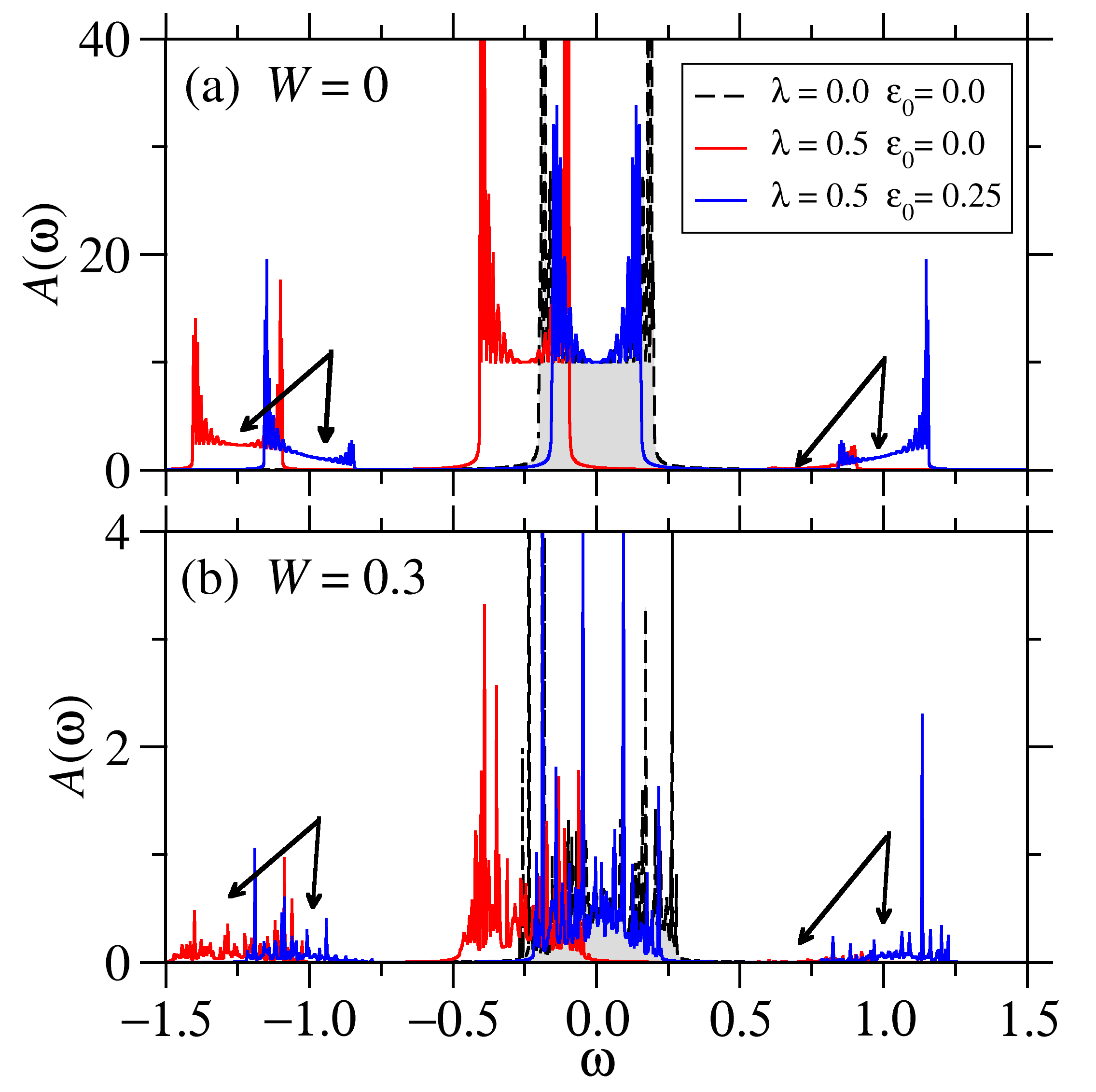}}
\caption{(Color online) Spectral function $A(\omega)$ at $k_{B}T=0.1$ for a MNW 
symmetrically coupled to leads ($\Gamma_{L}=\Gamma_{R}=0.2$) with (a)~no 
disorder and (b)~random on-site energies ($W=0.3$), averaged over $100$ 
realizations. The number of molecules is $N=20$. Blue and red solid lines 
correspond to finite electron-vibron coupling ($\lambda=0.5$) when 
$\varepsilon_{0}=0.25$ and $\varepsilon_{0}=0$, respectively. Black dashed lines 
show the spectral density of the non-interacting wire ($\lambda=0$) when 
$\varepsilon_{0}=0$ \revision{(the area under these two curves has been shaded to allow easier 
identification)}. For clarity, the error bars of the averages in panel~(b) 
are not shown. The arrows point at the first-order side 
bands of the interacting systems.}
\label{fig:spectral}
\end{figure}
%
In summary, the effect of the electron-vibron interaction on the spectral 
density of a uniform MNW is similar to the case of a single 
molecule~\cite{Chen05}. The most significant difference is the level splitting 
to form a band due to the intermolecular coupling and the resulting narrowing 
arising from the renormalization of the intermolecular hopping energy 
$\widetilde{J}$ with respect to the non-interacting case $J$.

We now introduce strong static disorder with magnitude $W=0.3$, which is of the order 
of the MNW bandwidth. This value is similar to those considered by Ciuchi and Fratini 
to discuss the temperature dependence of the mobility in rubrene organic field-effect
transistors~\cite{Ciuchi12}. It implies that the localization length in the
non-interacting lattice is smaller than the system size (strong disorder limit)
since it is determined from the ratio between the magnitude of disorder and the
bandwidth. In other words, this is a key parameter to elucidate the importance of 
disorder because the larger the ratio $W/J$, the smaller the localization 
length~\cite{Adame04}. When $W/J$ is of the order of unity the electron becomes mainly
localized at a single molecule. Figure~\ref{fig:spectral}(b) shows the resulting spectral 
function when the other parameters are the same as in Fig.~\ref{fig:spectral}(a). 
The displayed values were calculated by averaging over $100$ realizations. 
As expected, the obtained spectral densities show a much more complicated structure 
but the sum-rule $\int{A(\omega)\text{d}\omega}=2\pi$ remains valid, as the 
distribution of levels becomes random after introducing disorder. The effect of the 
electron-vibron interaction on the spectral function does not seem to differ 
much from the case without disorder. In particular, no midgap channels induced 
by disorder were found, in contrast to short molecular systems with a single 
defect.\cite{Nozaki10}

\section{Voltage-driven electric transport} \label{sec:J_dV}

We now turn to the impact of the electron-vibron interaction on the electric 
response of the MNW out of equilibrium. To this end, we calculate the non-linear 
dependence of the electric current given by Eq.~(\ref{eq:18a}) on the 
source-drain voltage $V$. As the charge transport across the wire is strongly 
dominated by resonant tunneling processes, the electric current and the 
differential conductance $\mathcal{G}(V,\Delta T=0)=dJ_{e}/dV\vert_{\Delta T=0}$ 
give a good insight into the complex non-linear transmission function of the 
system. We also calculate the heat current $J_{Q}$ from Eq.~\eqref{eq:18b}. We 
first discuss the simpler case of the MNW without disorder ($W=0$) and later 
compare it to the electric response of the system subjected to disorder 
($W=0.3$). In both cases the electric response is computed for a MNW with 
($\lambda=0.5$) and without ($\lambda=0$) electron-vibron interaction. The rest 
of parameters are the same as in Fig.~\ref{fig:spectral}.  

\subsection{Uniform molecular nanowires}

Figure~\ref{fig:GdE0} shows the non-linear conductance at~(a) a low temperature 
of $k_{B}T=0.01$ and~(b)  an intermediate temperature of 
$k_{B}T=0.2$ as a function of $eV$ for a MNW without disorder. The 
general shape of the low temperature conductance of the different cases in 
Fig.~\ref{fig:GdE0} follows the general trends of the spectral function shown in 
Fig.~\ref{fig:spectral}(a). This can be easily explained by the close relation 
between the conductance and the electronic transmission properties through the 
chain, with peaks corresponding to the resonant transmission channels shown by 
the spectral function. Notice the smearing out of the conductance due to the 
finite temperature and the symmetry of $\mathcal{G}(V,\Delta T=0)$ and 
$J_e(eV,\Delta T=0)$ against $\pm eV$. This is due to the upward and downward shift 
of the chemical potential of the left ($+eV/2$) and right ($-eV/2$) leads 
produced by the applied bias. Thus, a finite contribution of 
$\mathcal{G}(V,\Delta T=0)$ will correspond to energies $\omega=\pm eV/2$ where 
the spectral density is non-zero.

\begin{figure}[tb]
\centerline{\includegraphics[width=0.95\linewidth]{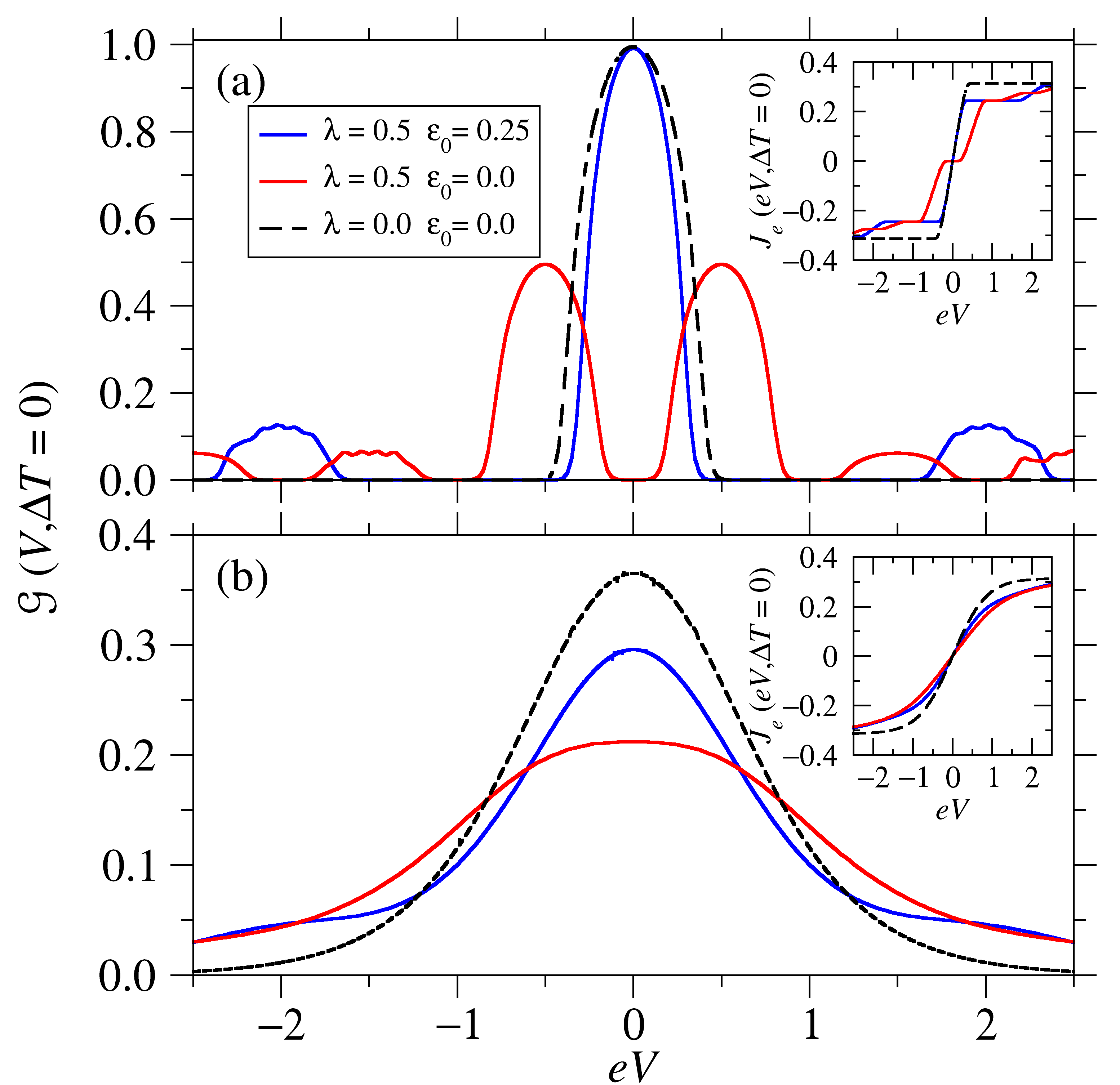}}
\caption{(Color online) Differential electric conductance 
$\mathcal{G}(V,\Delta T=0)=dJ_{e}/dV\vert_{\Delta T=0}$ as a function of $eV$ 
for a MNW without disorder at (a)~$k_{B}T=0.01$ and 
(b)~$k_{B}T=0.2$. Blue and red solid lines correspond to finite 
electron-vibron coupling ($\lambda=0.5$) when $\varepsilon_{0}=0.25$ and 
$\varepsilon_{0}=0$. Black dashed lines show the results for the non-interacting 
MNW ($\lambda=0$) when $\varepsilon_{0}=0$. The insets show the corresponding 
electric currents $J_e(eV,\Delta T=0)$. 
\label{fig:GdE0}}
\end{figure}

In the case without electron-vibron interaction shown in Fig.~\ref{fig:GdE0}(a) 
(black dashed line), the conductance in the uniform MNW only exhibits finite 
values within the band, as expected. For systems with finite electron-vibron 
interaction ($\lambda=0.5$) the differential electric conductance of a MNW with 
a renormalized on-site energy $\widetilde{\varepsilon}_{0}=0$ (blue solid line) 
resembles the case without interaction, with a maximum at $eV=0$ corresponding 
to the zero-vibron peak. The width of the peak is narrower due to the reduced 
hopping energy $\widetilde{J}<J$. Additionally, the differential 
conductance also shows side peaks at about $eV = \pm 2$, i.e., 
$\omega=\pm 1$, which match the transmission channels created by the absorption 
and emission of vibrons. When the renormalized on-site energy is nonzero (see 
red solid line, corresponding to $\widetilde{\varepsilon}_{0}=-0.25$), the 
central maximum disappears and two zero-vibron peaks at $eV=\pm 0.5$ arise 
instead. This is due to the fact that transmission channels are open only for 
either holes or electrons but not for both simultaneously. In addition, the red 
solid line shows two more side peaks at $eV=\pm 1.5$ and $eV=\pm 2.5$, the 
former matching the transmission channel at $\widetilde{\varepsilon}_{0}+1=0.75$ 
and the latter $\widetilde{\varepsilon}_{0}-1=-1.25$. Interestingly, the weight 
of the local maxima at $eV=\pm 1.5 $ is very large in comparison to the spectral 
function depicted in Fig.~\ref{fig:spectral}. The inset in 
Fig.~\ref{fig:GdE0}(a) shows the electric current $J_{e}(V,\Delta T=0)$ 
corresponding to the conductance of the main panel. $J_{e}(V,\Delta T=0)$ 
clearly displays the same saturation value independent of $\lambda$, which 
indicates that we do not induce any real scattering with the electron-vibron 
interaction and which corresponds to the conservation of the spectral sum-rule 
discussed in Sec.~\ref{sec:spectral}. Figure~\ref{fig:GdE0}(b) displays the 
conductance at a higher temperature $k_{B}T=0.2$. The different local 
maxima seen in Fig.~\ref{fig:GdE0}(a) reduce to a single wide peak due to the 
thermal smearing out of the Fermi-Dirac distribution in the leads. Also here 
the saturation values of the electric current with and without electron-vibron 
interaction are the same.

The MNW presents metallic or semiconducting transport properties, shown in 
Fig.~\ref{fig:GdE0}, according to the alignment of the states with respect to
the chemical potential. When the center of the band of states matches the chemical
potential in equilibrium $\mu=0$, the MNW is metallic and the current-voltage 
characteristics is linear around $eV=0$. This is the case of the 
non-interacting MNW when $\varepsilon_0=0$ [black dashed line in 
Figs.~\ref{fig:spectral} and~\ref{fig:GdE0}(a)], as seen in Fig.~\ref{fig:alignment}(a). 
The occurrence of a finite electron-vibron interaction shifts the band of states
and opens a gap, as depicted in Fig.~\ref{fig:alignment}(b). Consequently the MNW becomes
semiconducting. This is in agreement with the observation of a zero differential 
electric conductance at $eV=0$ shown in Fig.~\ref{fig:GdE0}(a) (red solid line). 
The small gap is not observed at high temperature, as expected [see 
Fig.~\ref{fig:GdE0}(b)]. Similar comments can be done regarding the side bands 
of the spectral function that also reveal themselves in the differential electric
conductance curves (not shown in Fig.~\ref{fig:alignment} for the sake of clarity).

\begin{figure}[tb]
\centerline{\includegraphics[width=0.8\linewidth]{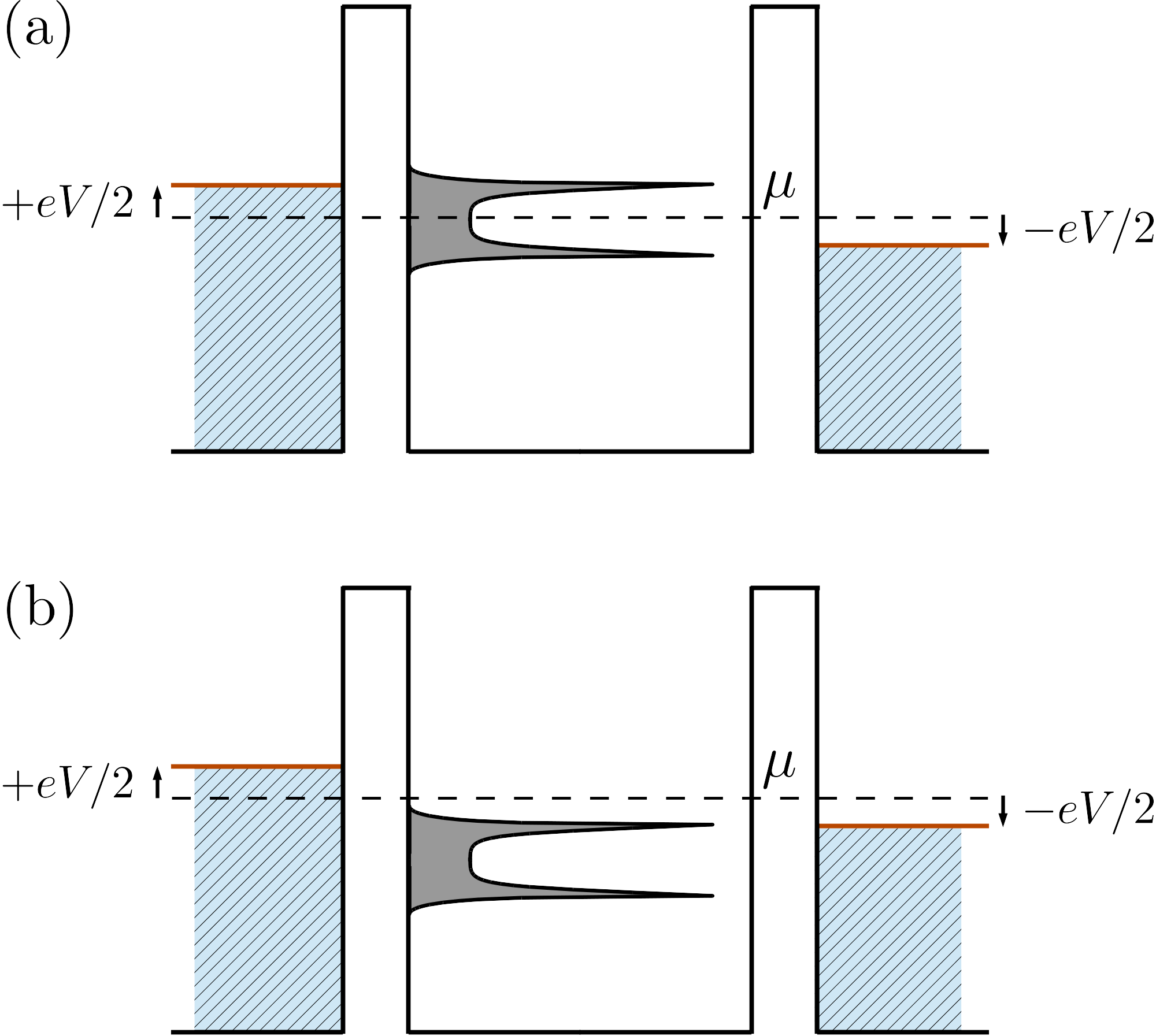}}
\caption{(Color online) Level alignment around the chemical potential $\mu$ when 
the MNW is subject to an applied voltage. (a)~When the band of states is located
symmetrically about $\mu$ the MNW presents metallic behavior [solid blue and black 
dashed lines in Figs.~\ref{fig:spectral} and~\ref{fig:GdE0}(a)]. (b)~A finite and
large value of the electron-vibron interaction shifts downward the band of states 
and a gap opens [red solid line in Fig.~\ref{fig:GdE0}(a)].}
\label{fig:alignment}
\end{figure}

Figure~\ref{fig:JqVdE0} shows the heat current $J_{Q}(V)$ as a function of $eV$ 
at low and intermediate temperatures as in Fig.~\ref{fig:GdE0}. Unlike the 
electric current, the heat current does not saturate at high $eV$ due to Joule 
heating. The observed linear behavior of $J_{Q}(V,\Delta T=0)$ at high voltage 
manifests itself in the saturation of the differential electrothermal 
conductance $\mathcal{M}(V,\Delta T=0)=dJ_{Q}/dV\vert_{\Delta T=0}$ (shown in 
the insets). The electron transmission properties can be depicted in the 
non-linear progression of the differential electrothermal conductance 
$\mathcal{M}$, where we see strong deviations compared to the differential 
electric conductance $\mathcal{G}$ due to the Joule heating $\mu_LJ_{e}/e$ and 
the weighting by the tunneling electron energy $\omega$ in Eq.~(\ref{eq:18b}). 
One should note the strong resemblance of the $J_{Q}(V,\Delta T=0)$ curves 
without electron-vibron interaction (black dashed lines) and the corresponding 
curve with the coupling switched on for the same bare on-site energy 
$\varepsilon_{0}=0$ (red solid lines) for $eV>0$. This can be explained by the 
small contribution of thermally generated vibrons at low background 
temperatures, so that the total energy is nearly conserved when switching on the 
electron-vibron interaction. Therefore, when the electric bias is high enough 
and all vibronic side bands are open for transmission, the total energy 
transferred to the system is approximately conserved after switching on the 
interaction. On the contrary, this conservation does not hold for an electric 
bias such $eV<0$ due to the breaking of the electron-hole symmetry (asymmetry of 
the transmission about $\omega=0$) by switching on the interaction, as discussed 
in the previous section. Unlike the symmetric electric current $J_{e}(V)$, this 
asymmetry transmission can be seen in the heat current from the left lead to the 
MNW~(\ref{eq:18b}). On the other hand, the blue solid lines, which correspond to 
a renormalized on-site energy $\widetilde{\varepsilon}_{0}=0$ and therefore a 
symmetric transmission, also shows a heat current symmetric about $eV=0$.

\begin{figure}[tb]
 \centerline{\includegraphics[width=0.95\linewidth]{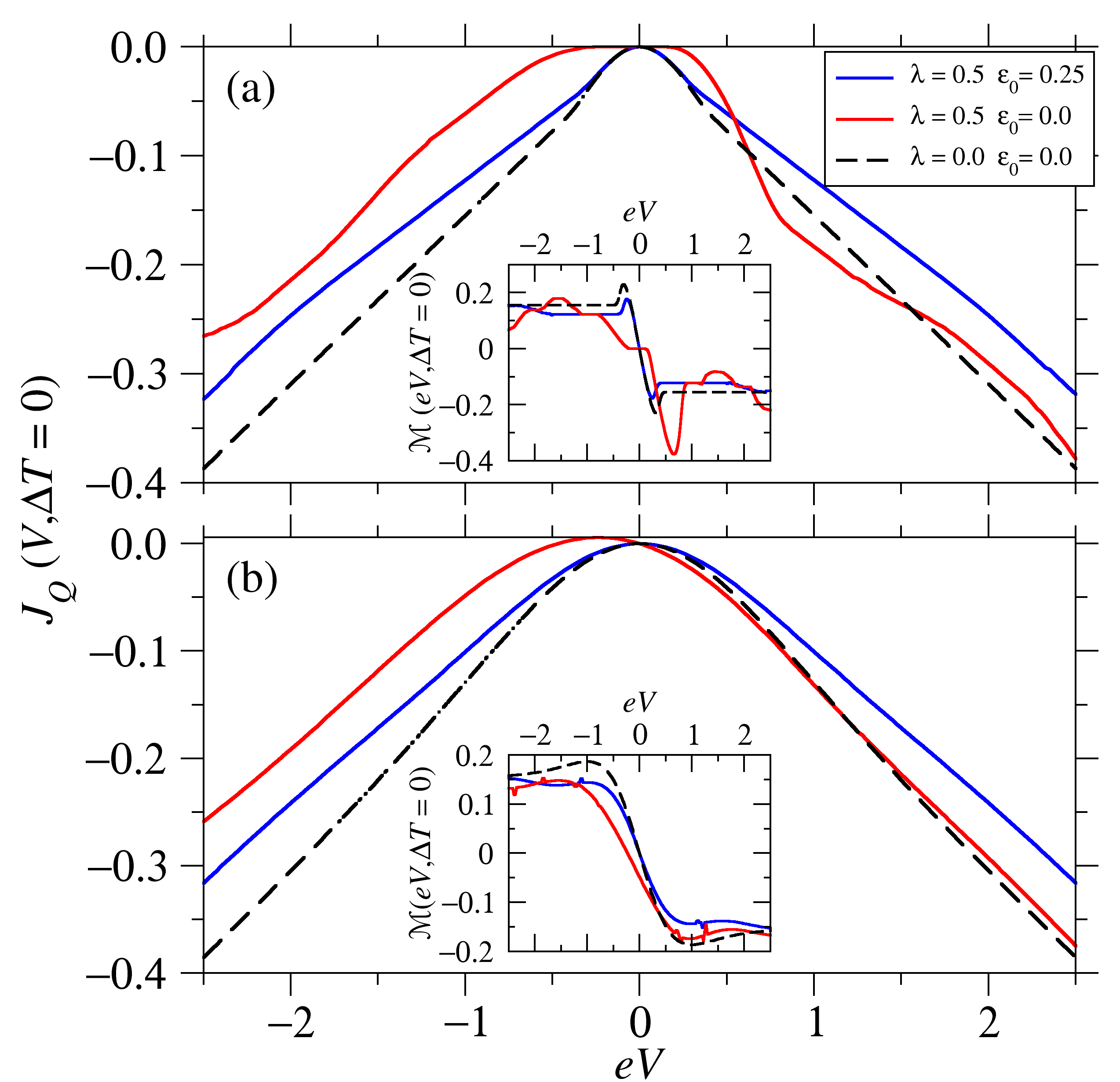}}
\caption{(Color online) Heat current $J_{Q}(V,\Delta T=0)$ as a function of $eV$ for MNWs without 
disorder as in the case of Fig.~\ref{fig:GdE0} at (a)~$k_{B}T=0.01$ and 
(b)~$k_{B}T=0.2$. Blue and red solid lines correspond to finite 
electron-vibron coupling ($\lambda=0.5$) when $\varepsilon_{0}=0.25$ and 
$\varepsilon_{0}=0$. Black dashed lines show the results for the non-interacting 
wire ($\lambda=0$) when $\varepsilon_{0}=0$. The insets display the 
corresponding differential electrothermal conductance $\mathcal{M}(V,\Delta 
T=0)=dJ_{Q}/dV\vert_{\Delta T=0}$.}
\label{fig:JqVdE0}
\end{figure}

\subsection{Disordered molecular nanowires}

Figure~\ref{fig:GdE0_3}(a) shows the differential electric conductance 
$\mathcal{G}(V,\Delta T=0)$ of disordered MNWs with $W=0.3$ at 
$k_{B}T=0.01$, averaged over $100$ realizations, and the other parameters 
as in Fig.~\ref{fig:GdE0}. The main features of the curves resemble those 
without disorder, the main difference being the occurrence of sharper peaks and 
a strong reduction of the maximum differential electric conductance. The 
conductance is reduced by a factor of $\sim 8$ for the chain without 
electron-vibron interaction, and up to $\sim 15$ for the interacting case. When 
temperature is increased to $k_{B}T=0.2$ the peaks of the 
differential conductance become broader and even merge into a single one in the 
case $\lambda=0.5$ and $\varepsilon_{0}=0$ (not shown) as for the ordered MNW 
in Fig.~\ref{fig:GdE0}(b). 

\begin{figure}[tb]
\centerline{\includegraphics[width=0.95\linewidth]{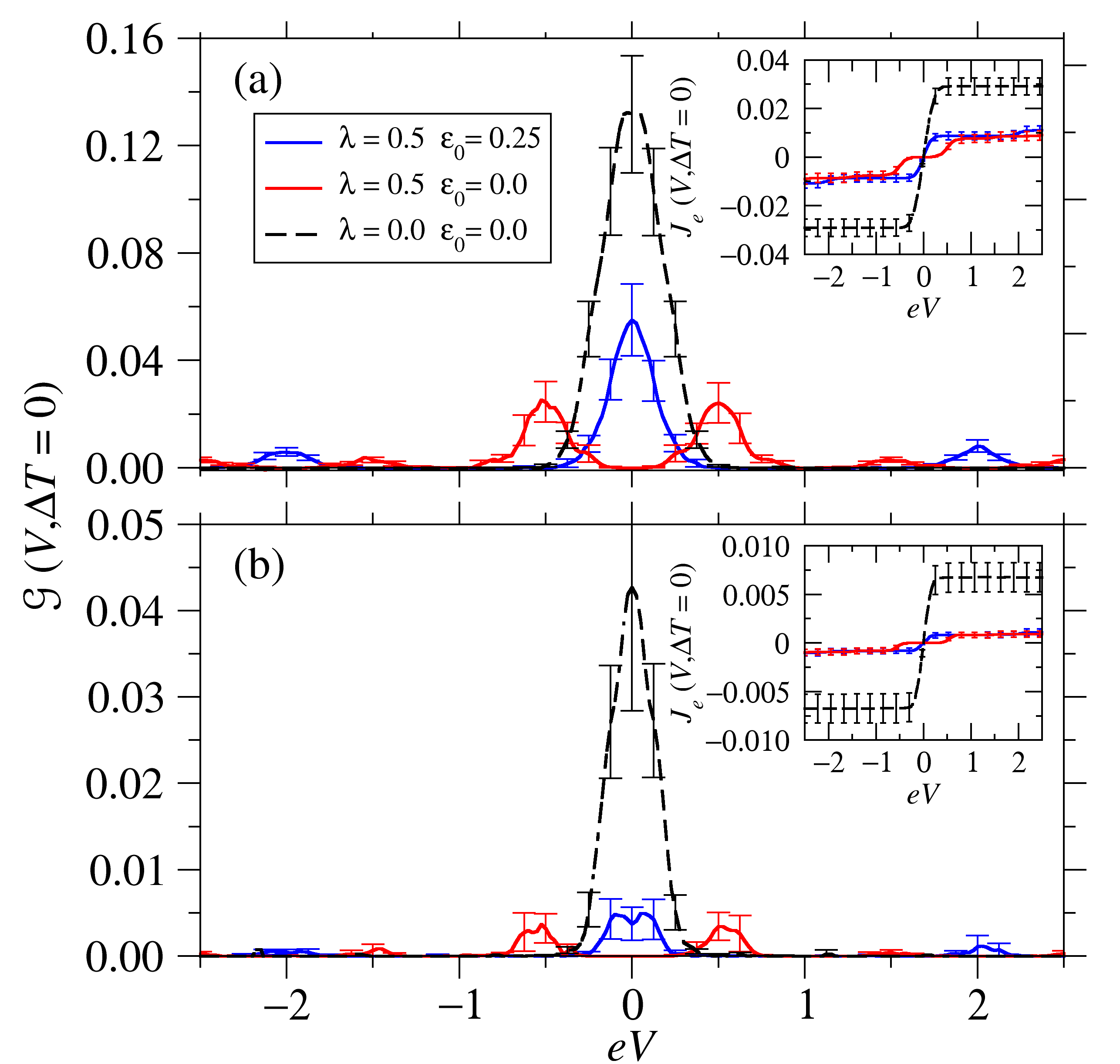}}
\caption{(Color online) Differential electric conductance 
$\mathcal{G}(V,\Delta T=0)=dJ_{e}/dV\vert_{\Delta T=0}$ at 
$k_{B}T=0.01$ as a function of $eV$ of MNWs of length (a)~$N=20$ and 
(b)~$N=40$ subjected to disorder ($W=0.3$). Results were averaged over $100$ 
realizations. The other parameters are the same as in Fig.~\ref{fig:GdE0}. Blue 
and red solid lines correspond to finite electron-vibron coupling 
($\lambda=0.5$) when $\varepsilon_{0}=0.25$ and $\varepsilon_{0}=0$. Black 
dashed lines show the results for the non-interacting wire ($\lambda=0$) when 
$\varepsilon_{0}=0$. The inset shows the corresponding electric current $J_e(eV,\Delta T=0)$.}
\label{fig:GdE0_3}
\end{figure}

In addition, the differential electric conductance is further decreased if the 
length of the MNW increases from $N=20$ to $N=40$, as displayed in 
Fig.~\ref{fig:GdE0_3}(b). This behavior is in contrast to the uniform MNW, where 
the conductance is preserved against the system length. It is worth noting the 
stronger reduction in the MNW with finite electron-vibron interaction when 
doubling the length, compared to the case without interaction. Such a reduction 
can be traced back to the Anderson-localization of electron states 
\cite{Anderson58,Mott61,Abrahams79}. In general, the magnitude of disorder has 
to be compared with the hopping energy or, in other words, with the bandwidth of 
the uniform system. The larger the ratio between them, the smaller the 
localization length. From this reasoning, the effects of disorder should be more 
important on increasing temperature when the electron-vibron coupling is finite 
since the exponential suppression of tunneling reduces the dressed 
intermolecular hopping energy $\widetilde{J}$. The expected reduction of the 
localization length is hinted from the comparison of Figs.~\ref{fig:GdE0} and 
\ref{fig:GdE0_3}.    

For completeness, Fig.~\ref{fig:JqVdE0_3} shows the \revision{much reduced} heat current 
$J_{Q}(V,\Delta T)$ in disordered MNWs with $W=0.3$ as compared to 
Fig.~\ref{fig:JqVdE0}. The influence of disorder agrees well with the 
observed electric current, supporting the idea of an increased localization in 
the case of finite electron-vibron interaction that influences both charge and 
heat transport alike.

\begin{figure}[tb]
\centerline{\includegraphics[width=0.95\linewidth]{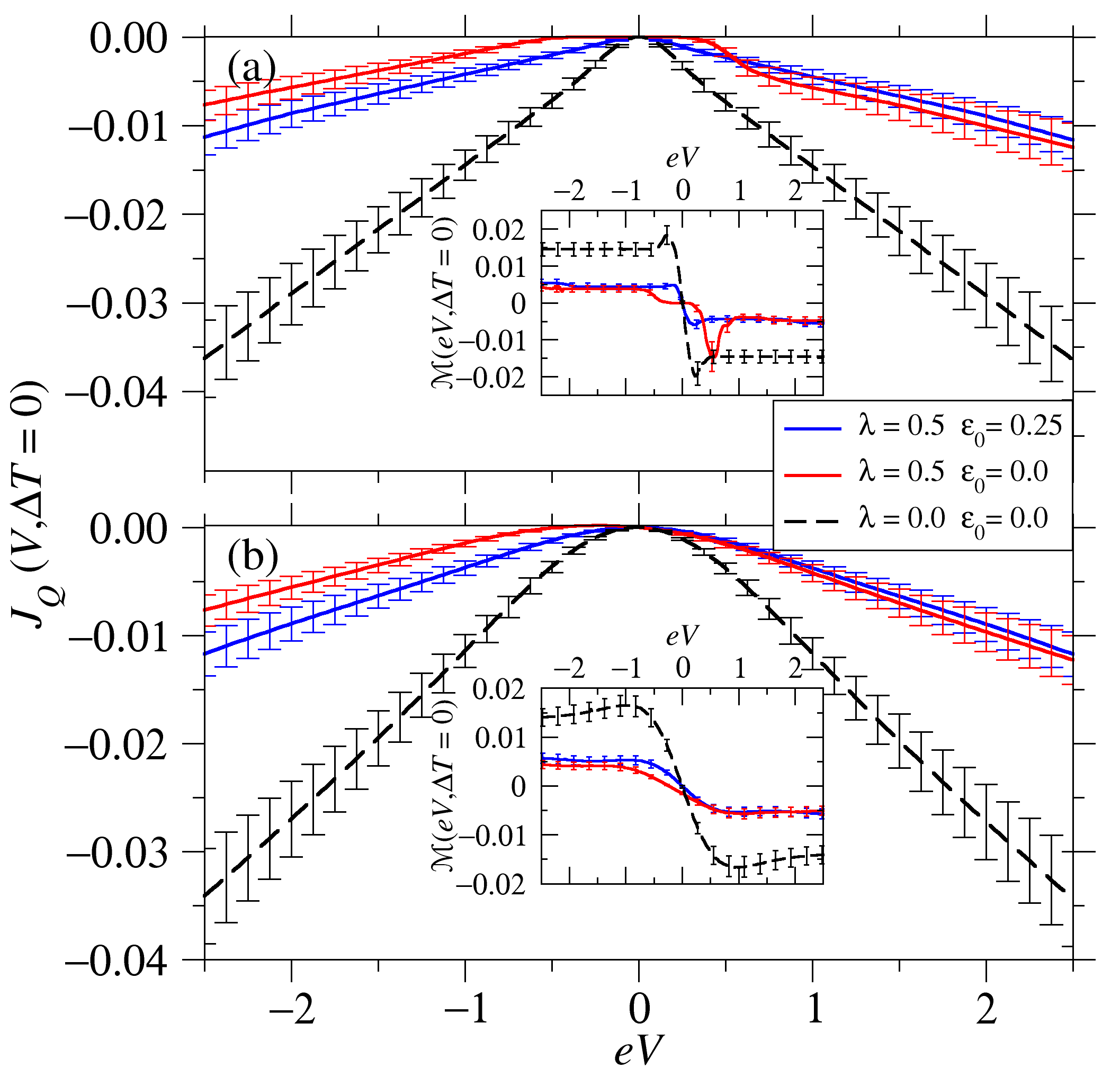}}
\caption{(Color online) Heat current $J_{Q}(V,\Delta T=0)$ as a function of $eV$ of MNWs with disorder 
($W=0.3$) averaged over $100$ realisations at (a)~$k_{B}T=0.01$ and 
(b)~$k_{B}T=0.2$. The other parameters are the same as in 
Fig.~\ref{fig:JqVdE0}. Blue and red solid lines correspond to finite 
electron-vibron coupling ($\lambda=0.5$) when $\varepsilon_{0}=0.25$ and 
$\varepsilon_{0}=0$. Black dashed lines show the results for the non-interacting 
wire ($\lambda=0$) when $\varepsilon_{0}=0$. The insets show the corresponding 
differential electrothermal conductance $\mathcal{M}(V,\Delta T=0)=dJ_{Q}/dV\vert_{\Delta T=0}$.} 
\label{fig:JqVdE0_3}
\end{figure}

\section{Temperature-driven electric transport}

We now investigate the electric transport through MNWs in response to a 
temperature difference $\Delta T$ only. We are assuming a symmetrically biased 
system with $T_L=T-\Delta T/2$ and $T_R=T+\Delta T/2$, therefore keeping the 
average temperature $(T_{L}+T_{R})/2$ at a defined value $T$. In order to ensure 
that each site-dependent $T_{i}>0$, the maximum $\Delta T_\mathrm{max}$ is fixed 
at $\Delta T_\mathrm{max}=\pm T/2$ so that the leads do not deviate more then 
$25~\%$ from their initial temperature. In this section, the bias voltage is 
absent ($V=0$) and charge flows through the MNW only due to the temperature 
difference between the leads. As in Sec.~\ref{sec:J_dV} we first discuss the 
electric and heat currents in uniform MNWs ($W=0$) and later we consider random 
on-site energies with $W=0.3$. The other parameters of the system are taken the 
same as in Sec.~\ref{sec:J_dV}, with the additional case of a non-interacting 
MNW with on-site energy $\varepsilon_{0}=0.25$.

\subsection{Uniform molecular nanowires}

Figure~\ref{fig:Je+Jq_dTdE0}(a) shows the electric current $J_{e}(V=0,\Delta T)$ 
at $k_{B}T=0.2$ as a function of $\Delta T$ for a MNW without disorder. 
The blue solid line ($\lambda=0.5$ and $\widetilde{\varepsilon}_{0}=0$) and the 
black dashed line ($\lambda=0$ and $\varepsilon_{0}=0$), which represent systems 
with symmetric transmission around the Fermi energy $\mu_L=\mu_R=0$, do not 
exhibit any thermoelectric current at all. This is due to the fact that, in case 
of a thermal bias, all electrons tunneling from the hot lead to the unoccupied 
states of the cold lead are compensated by holes tunneling from the unoccupied 
states to the still occupied state of the cold lead due to the electron-hole 
symmetry. If this symmetry is broken, as in the case of the red solid line 
($\lambda=0.5$ and $\widetilde{\varepsilon}_{0}=-0.25$) and the blue dotted line 
($\lambda=0$ and $\varepsilon_{0}=0.25$), the system exhibits a finite 
thermally-driven electron or hole current. 
The inset of Fig.~\ref{fig:Je+Jq_dTdE0}(a) links these results to those related to the 
spectral density in Sec.~\ref{sec:spectral}.
According to Eq.~(\ref{eq:18a}) the most relevant contribution to $J_e$ can be approximated by the frequency integration of the magnitude $[f_L(\omega)-f_R(\omega)]\mathcal{A}(\omega)$. The inset plots the odd function $f_L(\omega)-f_R(\omega)$ together with colored areas representing the frequency range where the spectral densities are relevant. As seen in Fig.~\ref{fig:spectral}(a) the spectral density is centered at $\omega=0$ when $\widetilde{\varepsilon}_{0}=0.0$ (blue area) and it is shifted to lower (higher) frequencies when $\widetilde{\varepsilon}_{0} <0.0$ ($\widetilde{\varepsilon}_{0} > 0.0$) (red and black areas). Thus, it is clear that the signs and symmetries of both factors justify the values of $J_e$ presented in the main plot.
Unlike the voltage-driven case, this 
current is nearly linear since the temperature difference is not high enough to 
reach a non-linear regime. 

\begin{figure}[tb]
\centerline{\includegraphics[width=0.95\linewidth]{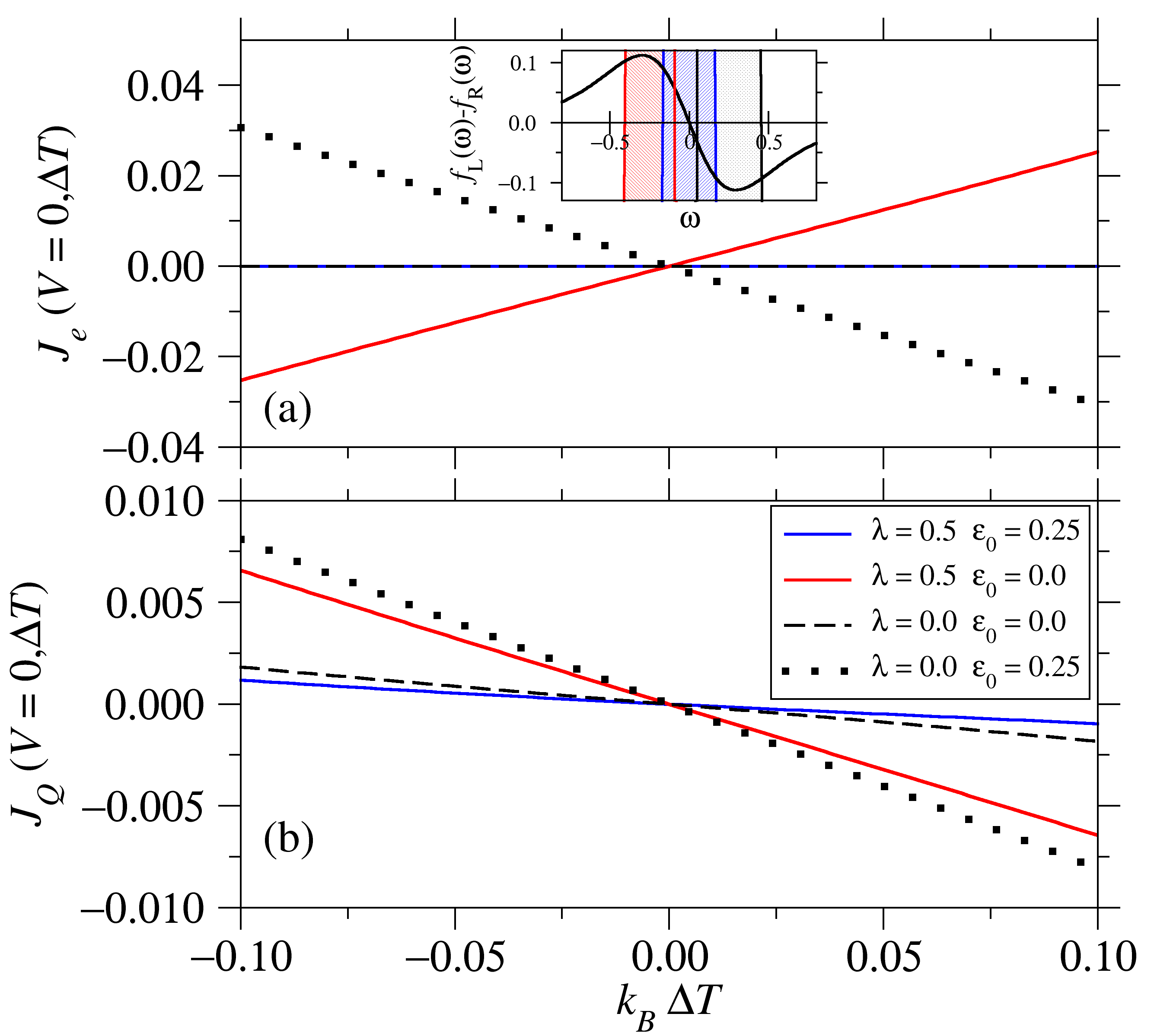}}
\caption{(Color online) Temperature-driven (a)~electric current $J_{e}$ and 
(b)~heat current $J_Q$ as a function of the temperature difference $\Delta T$ 
for MNWs without disorder ($W=0$) at $k_{B}T=0.2$. Blue and red solid 
lines correspond to finite electron-vibron coupling ($\lambda=0.5$) when 
$\varepsilon_{0}=0.25$ and $\varepsilon_{0}=0$. Black dashed (superimposed on the
blue solid line) and dotted lines show the results for the non-interacting wire 
($\lambda=0$) when $\varepsilon_{0}=0$ and $\varepsilon_{0}=0.25$, respectively. Inset represents the function $f_L(\omega)-f_R(\omega)$ as a function of the frequency $\omega$ together with a colored area representing the frequency region where the spectral density is finite for every parameter set. }
\label{fig:Je+Jq_dTdE0}
\end{figure}

Figure~\ref{fig:Je+Jq_dTdE0}(b) displays the heat current $J_{Q}(V=0,\Delta T)$ 
at $k_{B}T=0.2$ as a function of $\Delta T$. Unlike the electric current 
$J_{e}(V=0,\Delta T)$, also the blue solid line ($\lambda=0.5$ and 
$\widetilde{\varepsilon}_{0}=0$) and the black dashed line ($\lambda=0$ and 
$\varepsilon_{0}=0$) exhibit finite values due to weighting of the tunneling 
particles with their respective energies and the asymmetry of the heat currents 
from the left and the right leads ($J_{Q}^{L}\neq J_{Q}^{R}$). As the majority 
of the carriers are transmitted through the zero-vibron channel, the transmitted 
energy and therefore the heat current is rather low compared to the other two 
cases, whose main transmission channels are centered at a finite energy. A small 
influence of the electron-vibron interaction can be seen as a slight asymmetry 
of the absolute value of the thermal current $\left|J_{Q}(\Delta T)\right| \neq 
\left|J_{Q}(-\Delta T)\right|$ for the cases with finite $\lambda$. Such an 
effect can also be depicted for disordered MNWs in the next section and it shall 
be discussed on that behalf.
 
\subsection{Disordered molecular nanowires}

Figure~\ref{fig:Je+Jq_dTdE0_3}(a) shows the average temperature-driven $J_{e}(V=0,\Delta T)$ as a function of 
$\Delta T$ for the same systems shown in Fig.~\ref{fig:Je+Jq_dTdE0} but with 
disorder $W=0.3$. As in the case of voltage-driven transport 
discussed in Sec.~\ref{sec:J_dV}, disorder strongly alters the electric response 
of the MNW. We observe that it affects the interacting MNW more significantly 
than the non-interacting one. This can be explained by the same reasoning 
introduced in Sec.~\ref{sec:J_dV}.

\begin{figure}[ht]
\centerline{\includegraphics[width=0.95\linewidth]{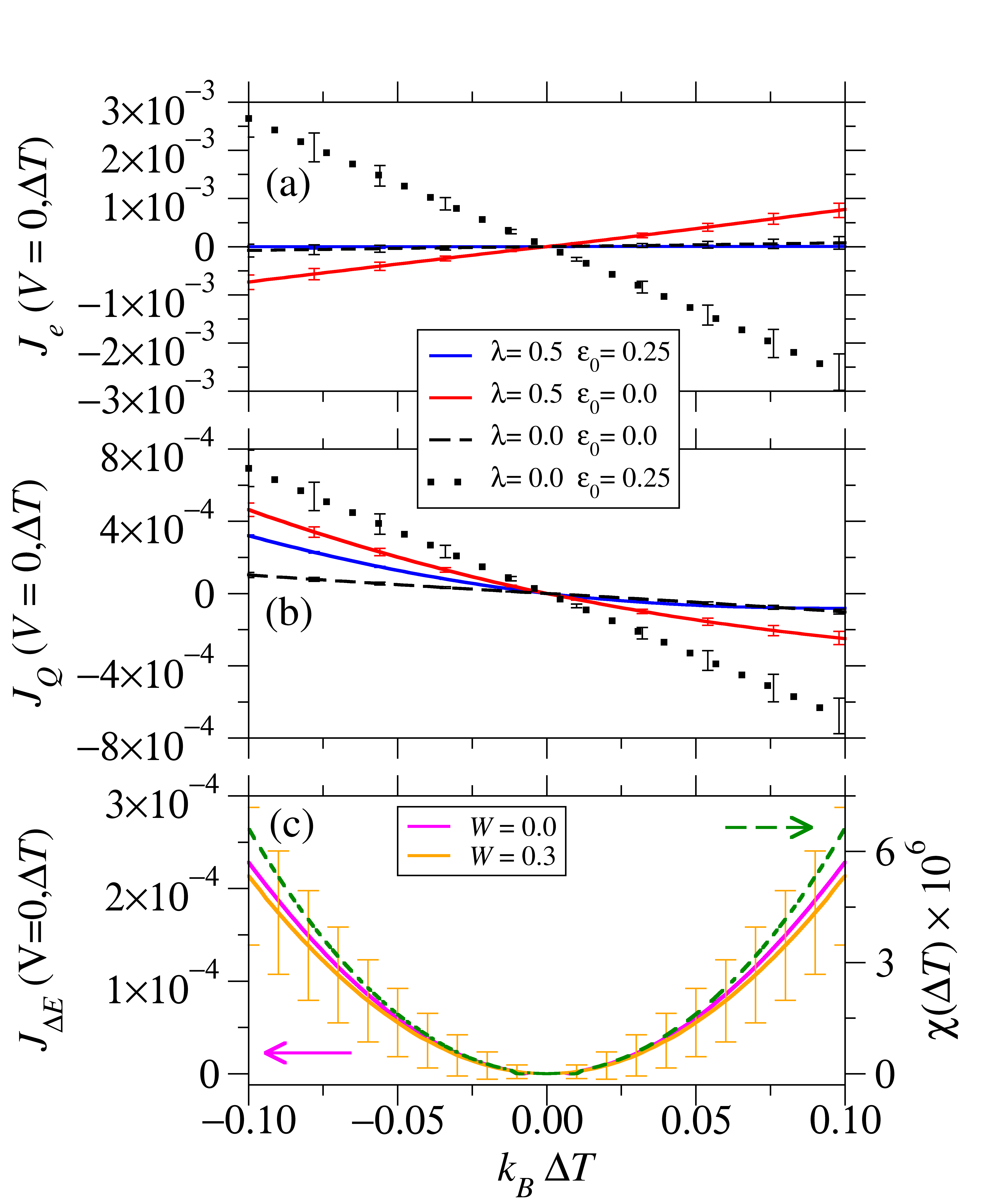}} 
\caption{(Color online) Temperature-driven (a)~electric current $J_{e}$ and  
(b)~heat current $J_Q$  as a function of the temperature difference $\Delta T$ 
for MNWs with disorder ($W=0.3$) at $k_{B}T=0.2$. Blue and red solid 
lines correspond to finite electron-vibron coupling ($\lambda=0.5$) when 
$\varepsilon_{0}=0.25$ and $\varepsilon_{0}=0$. Black dashed and dotted lines 
show the results for the non-interacting wire ($\lambda=0$) when 
$\varepsilon_{0}=0$ and $\varepsilon_{0}=0.25$, respectively. Panel~(c) displays the  
heat generation inside the MNW $J_{\Delta E}$ as a function of $\Delta T$ at 
$k_{B}T=0.2$ for an $\varepsilon_{0}=0.25$ in the case without disorder 
(magenta solid line) and with disorder (orange solid line) calculated from 
Fig.~\ref{fig:Je+Jq_dTdE0}(b) and~\ref{fig:Je+Jq_dTdE0_3}(b) respectively. The 
green dashed line represents the occupation number of thermally generated vibrons times 
the thermoelectric current $\chi(\Delta T)$ in arbitrary units as a function of $\Delta T$.}
\label{fig:Je+Jq_dTdE0_3}
\end{figure}

Figure~\ref{fig:Je+Jq_dTdE0_3}(b) displays the temperature-driven  
$J_{Q}(V=0,\Delta T)$ for the same parameters as in 
Fig.~\ref{fig:Je+Jq_dTdE0_3}(a). One can see a strong non-linear curve 
progression for MNWs with finite coupling. More importantly, the interacting 
system actually exhibits a stronger or nearly equal heat current in comparison 
with the non-interacting one, at least for $\Delta T<0$. As can be concluded 
from the effect of disorder seen in the electric current in 
Fig.~\ref{fig:Je+Jq_dTdE0_3}(a), this can not be related to the electric 
response of the system. Therefore, it must arise from the heat transport 
properties. Due to the symmetry of the system 
at $\Delta T=0$, $J_{Q}^{L}(-\Delta T)=J_{Q}^{R}(\Delta T)$, the rate of energy 
generation inside the system is $J_{\Delta E}(\Delta T)
\equiv J_{Q}^{L}(\Delta T)+J_{Q}^{L}(-\Delta T)$. Thus, the asymmetry 
between the $J_{Q}^{L}(-\Delta T)$ and $J_{Q}^{L}(\Delta T)$ must be explained 
by a heat generation process inside the system. As this heat generation process 
is only noticeable in interacting MNWs, it is reasonable to assume that it is 
due to the thermal generation of vibrons. At $k_{B}T=0.2$ the amount of 
thermally generated vibrons is very small, which is why the asymmetry is much 
less visible in case of no disorder in Fig.~\ref{fig:Je+Jq_dTdE0}(b) due to an 
order of magnitude higher overall $J_{Q}^{L}(\Delta T)$. But when actually 
calculating the heat generation rate $J_{\Delta E}$, as depicted in 
Fig.~\ref{fig:Je+Jq_dTdE0_3}(c) for the case of $\varepsilon_0=0$, in good 
approximation one gets the same values for the ordered (magenta curve) and the 
disordered MNW (red curve). This justifies the explanation of an effect governed 
by the vibronic subsystem. In addition, the effect of heat generation gets 
further support when looking at the green curve in 
Fig.~\ref{fig:Je+Jq_dTdE0_3}(c) that shows the magnitude
$\chi(\Delta T)=J_e(V=0,\Delta T)\sum_{i=1}^{N} n_i(T_i)$. It represents occupation 
number of thermally generated vibrons inside the system times the electric 
current and, consequently, it provides a rough estimation of the number of electrons 
to which the vibrons can couple. As the green curve clearly is proportional 
to the heat generation rate $J_{\Delta E}$, one can safely conclude that the additional 
heat is transferred from the heat baths to the MNW.  

\section{Conclusions}

In conclusion, we have studied the non-equilibrium transport properties of a 
disordered MNW. The wire is regarded as a quasi-one-dimensional organic crystal 
of random single-level molecules, connected in series to two leads. We also have 
assumed that the electron interacts with local vibrational modes in the 
molecules and investigated the effects of the interaction on the electric and 
heat currents in response to either an applied electric bias or temperature 
gradient established in the system. The original many-body problem has been 
turned into an effective one-body problem by the polaron (Lang-Firsov) transformation. 

We have considered the regime of strong disorder, for which the localization 
length in the non-interacting MNW ($\lambda=0$) is smaller than the system size. In 
addition, we have taken $k_{B}T\ll \omega_0$ in our simulations and 
consequently the scattering is mainly dominated by the interaction with the 
disordered lattice. This is supported by the fact that the voltage-driven 
electric current $J_e$ in uniform MNWs ($W=0$) is independent of the system 
length and its saturation value is the same for both non-interacting and 
interacting cases. \revision{In general, voltage- and temperature-driven electric currents
present a similar decrease due to disorder or electron-vibron coupling}. 
Remarkably, we have found that the electron-vibron 
interaction enhances the effects of the disorder on the electric and heat 
currents. This important result can be understood as follows. The intermolecular 
hopping energy $\widetilde{J}_i$ in the interacting MNW is smaller than the bare 
$J$ due to the occurrence of the exponential suppression of 
tunneling~\cite{Rudzinski08}. Therefore, disorder is effectively stronger when 
the interaction is switched on because the ratio between the magnitude of 
disorder and the bandwidth increases.

Regarding the temperature-driven transport, we have numerically found an almost 
linear dependence of the electric and heat currents on the temperature 
difference. In MNWs with preserved electron-hole symmetry the electric current 
vanishes even if the electron-vibron interaction is taken into account. On the 
contrary, the heat current is always non-zero and enhanced due to the 
interaction, although to a small extent. Most importantly, the effects of 
disorder are more pronounced in the interacting MNW. \revision{Disorder reveals
itself by a slight deviation of the temperature-driven heat current curves from linearity.
In addition, temperature-driven heat current is less sensitive to the increase of 
the magnitude of disorder or the electron-vibron coupling than the electric current.}

In order to get a clear connection with experiments, a look at the magnitudes of interest 
in physical units is in order. Our energy unit thus far has been the vibron frequency 
$\omega_0$ so the following numbers depend strongly on the details of our molecular bridge (see Ref.~\onlinecite{Galperin07} for a review on this topic). By taking a reference value of 
$\omega_0= 150\,$meV, our study focuses on a physical scenario where the conducting molecular 
level is $\epsilon_0 \sim 40-80\,$meV and the electron coupling with the vibrons and the 
leads are $\lambda=10\,$fs and $\Gamma^{L/R}=30\,$meV, respectively. These parameters 
are consistent with those found by experiments~\cite{Poot06} with charge and heat 
currents of the order of $J_e\sim 10-100\,$nA and $J_Q\sim 1-10\,$nW  at achievable 
temperatures of $T\sim 15-350\,$K.

Although this work is focused on the study to MNWs, our results can be extended 
to arrays of quantum dots as well. These arrays can be realized in electron 
gases with superposed mesh gates. Beside their interest in fundamental research, 
they are regarded as good candidates for building quantum 
simulators.\cite{Buluta09,Chorley08,Gray16} Unavoidable imperfections introduced 
during the fabrication process might have an impact on charge and energy 
transport when electrons are coupled to bosonic degrees of freedom. Our results 
shed light on the influence of disorder on the performance of quantum simulators 
based on arrays of quantum dots.

\acknowledgments

The authors are grateful to D. S\'{a}nchez, M.\ A.\ Sierra and C.~\'{A}lvarez 
for helpful discussions. F.~D-A. thanks the Theoretical Physics Group of the 
University of Warwick for the warm hospitality. Work at Madrid has been 
supported by MINECO under Grants MAT2013-46308 and MAT2016-75955. UK research data 
statement: All data accompanying this publication are directly available within the 
publication.

\appendix

\section{Lang-Firsov polaron transformation  \label{sec:appA}}

Starting from Eqs.(\ref{eq:01}-~\ref{eq:03}), we apply the polaron (Lang-Firsov~\cite{Lang63})
nonperturbative 
canonical transformation $\widetilde{H}=e^S H e^{-S}$, where the operator $S$ is defined as 
$S=(\lambda/\omega_0)\sum_i  (a_{i}^{\dagger}-a_{i}^{})c_{i}^{\dagger}c_{i}^{}$~\cite{Lang63}. 
This transformation yields the following transformed Hamiltonian
\begin{align}
\widetilde{H}&=\sum_{i=1}^{N}\widetilde{\varepsilon}_i^{}
c_i^{\dagger}c_i^{} - J\sum_{i=1}^{N-1} 
\left(X_{i}^{\dag}X_{i+1}^{}c_{i}^{\dagger}c_{i+1}^{} 
+ \mathrm{H.c}\right)
\nonumber \\
&+\omega_{0} \sum_{i=1}^{N} a_i^{\dagger}a_i^{} 
+\sum_{\alpha k i} \left(V_{\alpha k i}^{}X_{i}^{}
d_{\alpha k}^{\dagger}c_i^{} + \mathrm{H.c}\right) \ ,
\label{eq:04}
\end{align}
where $\widetilde{\varepsilon}_i=\varepsilon_i-\lambda^2 /\omega_0$ is the 
renormalized energy level of the molecule~$i$. After the transformation, a new 
operator arises, $X_i=\exp[-(\lambda/\omega_0)(a_i^{\dagger}-a_i^{})]$, named 
the displacement operator. The canonical transformation is exact but it does not 
diagonalize the Hamiltonian. In other words, $\widetilde{H}$ still contains 
products of boson and fermion operators. When the coupling to the leads is weak 
($|V_{\alpha k i}| < \lambda$), it is reasonable to replace the displacement 
operator $X_i$ by its thermal expectation value evaluated in equilibrium 
$\langle X_i \rangle=\exp\left[-\xi_i(T_i)/2\right]$~\cite{Chen05,Wu12}, where 
$\xi_i(T_i)=g\left(2n_{i}+1\right)$,
$n_{i}=1/\left[\exp\left(\omega_0/k_{B}T_i\right)-1\right]$
and the Huang-Rhys factor is $g=\lambda^2/\omega_0^2$. 
By way of this procedure one can deal with an effective one-body problem according to 
Eq.~(\ref{eq:05}). 

\section{Non-equilibrium Green's functions  \label{sec:appB}}

The various Green's functions of the system described by Eq.(\ref{eq:05}) are lengthy 
but straightforward to calculate numerically with the help of the Keldysh non-equilibrium 
Green's function formalism~\cite{Keldysh65,Galperin07}. First, since we replaced the 
displacement operator $X_i$ by its thermal expectation value, the Green's functions 
are then factored out. The greater Green's function can be cast in the form
\begin{equation}
G_{ij}^{>}(t)=  
\widetilde{G}_{ij}^{>}(t)\langle X_i^{}(t) X_j^{\dagger}(0) \rangle\ ,
\label{eq:06}
\end{equation}
where $\widetilde{G}_{ij}^{>}(t)$ denotes the so called dressed greater Green's 
function. $i\widetilde{G}_{ij}^{>}(t)$ is the correlation function of a hole
dressed by vibrons. Concerning the vibron part, we may encounter two different 
cases. If $i\neq j$ then $\langle X_i(t) X_j^{\dagger}(0) \rangle 
=\exp\left[-\xi_i(T_i)/2-\xi_j(T_j)/2\right]$. When $i=j$ the calculation 
of the correlation function is more involved but can be performed analytically. 
The details are presented in Ref.~\onlinecite{Mahan00} and the final result is
%
\begin{equation}
\langle X_i(t) X_i^{\dagger}(0) \rangle =
\sum_{n=-\infty}^{\infty} L_n^i(T_i) e^{\rmi n\omega_0 t} \ ,
\label{eq:07a}
\end{equation}
where at finite temperature
\begin{equation}
L_n^i(T_i) = e^{-\xi_i(T_i)+n\omega_0/2k_{B}T_{i}}
I_n\left(\sqrt{\xi_i^2(T_i)-g^2}\,\right)\ ,
\label{eq:07b}
\end{equation}
with $I_n(z)$ the modified Bessel function of integer order, and at $T_i=0$
\begin{eqnarray}
L_n^i(0) = \begin{cases}
e^{-g}g^{n}/n! & \mathrm{if\ } n\ge 0\ ,  \\ 
0 & \mathrm{if\ } n < 0\ .
\end{cases}
\label{eq:07c}
\end{eqnarray}

Using the vibron mean values 
given above, the elements of the greater Green's functions, $G^>(\omega)$, are given in Fourier space as~\cite{Chen05}
\begin{subequations}
\begin{eqnarray}
G^{>}_{ii}(\omega)&=\sum_{n=-\infty}^{\infty} L_n^i(T_i)
\widetilde{G}^{>}_{ii}(\omega - \omega_0 n) \ ,
\\
G^{>}_{ij}(\omega)&=\langle X_i \rangle \langle X_j \rangle
\widetilde{G}^{>}_{ij}(\omega) \ , \qquad i \neq j\ ,
\label{eq:09}
\end{eqnarray}
\end{subequations}
and similarly for the lesser Green's function, $G^<(\omega)$, but replacing $\omega-\omega_0 n$ 
by $\omega+\omega_0 n$ in the summation. The dressed lesser and greater Green's 
functions can be calculated from the Keldysh equation 
$\widetilde{G}^{<(>)}(\omega)=\widetilde{G}^\mathrm{r}(\omega) 
\widetilde{\Sigma}^{<(>)}(\omega)\widetilde{G}^\mathrm{a}(\omega)$, where the 
self-energies are given by
\begin{subequations}
\begin{eqnarray}
\widetilde{\Sigma}^{<}(\omega)&=\rmi 
\left[f^{(e)}_L(\omega)\widetilde{\Gamma}^L+f^{(e)}_R(\omega)
\widetilde{\Gamma}^R\right]\ ,
\\
\widetilde{\Sigma}^{>}(\omega)&=-\rmi
\left[f^{(h)}_L(\omega)\widetilde{\Gamma}^L+f^{(h)}_R(\omega)
\widetilde{\Gamma}^R\right]\ ,
\label{eq:10}
\end{eqnarray}
\end{subequations}
with $f^{(e)}_\alpha(\omega)=f_\alpha(\omega)$ and 
$f^{(h)}_\alpha(\omega)=1-f_\alpha(\omega)$. Here 
$f_\alpha(\omega)=1/\left\{\exp\left[(\omega-\mu_\alpha)/k_{B}T_{\alpha}
\right] +1 \right\}$ is the Fermi-Dirac distribution function of the lead 
$\alpha$. The matrix elements of $\widetilde{\Gamma}^{\alpha}$ in~(\ref{eq:10}) 
are given as $\widetilde{\Gamma}^{\alpha}_{ij}=2\pi \rho_{\alpha} 
\widetilde{V}_{\alpha k i}\widetilde{V}^{*}_{\alpha k j}$, where $\rho_{\alpha}$ 
is the density of states of the corresponding lead. Notice that we will neglect 
their $k$-dependence by relying on the wide-band limit approximation and take 
$\widetilde{\Gamma}^{\alpha}$ matrices as energy-independent magnitudes. 

In order to calculate the dressed retarded Green's function $\widetilde{G}^\mathrm{r}(\omega)$, the 
equation-of-motion method is used. We start by calculating the time-derivative of its formal definition 
$\widetilde{G}^\mathrm{r}_{ij}(t)= -\rmi \theta(t)\langle \{c_i(t),c_j^{\dagger}(0)\}\rangle$ where 
$t \rightarrow t + \rmi 0^{+}$, 
keeping in mind that $\rmi \delta_t c_i(t)= [c_i(t),\widetilde{H}]$ 
where $\widetilde{H}$ is given by Eq.~(\ref{eq:05}) and $\theta(t)$ the Heaviside step function. 
After Fourier transform one can write the following system of linear equations to be solved
\begin{eqnarray}
(\omega-\widetilde{\varepsilon_i})\widetilde{G}^\mathrm{r}_{ij}(\omega)
&=&\delta_{ij}-\widetilde{J}_{i}\widetilde{G}^\mathrm{r}_{i+1,j}(\omega)
-\widetilde{J}_{i-1}\widetilde{G}^\mathrm{r}_{i-1,j}(\omega)\nonumber \\
&+&\sum_{l=1}^{N}\widetilde{\Sigma}_{il}^\mathrm{r}
\widetilde{G}^\mathrm{r}_{lj}(\omega)\ ,
\label{eq:16}
\end{eqnarray}
where
\begin{subequations}
\begin{equation}
\widetilde{\Sigma}_{il}^\mathrm{r}=\sum_{\alpha k} 
\frac{\widetilde{V}^{*}_{\alpha k i}\widetilde{V}_{\alpha k l}}%
{\omega-\varepsilon_{\alpha,k}+ \rmi 0^{+}}
\label{eq:17a}
\end{equation}
is the retarded self-energy. Within the wide-band approximation this term is 
written as
\begin{equation} 
\widetilde{\Sigma}_{il}^{r}(\omega)=\frac{\rmi}{2} 
\langle X_i \rangle \langle X_l\rangle (\Gamma_{il}^L+\Gamma_{il}^R)\ .
\label{eq:17b}
\end{equation}
Similarly, to calculate the dressed advanced Green's function $\widetilde{G}^\mathrm{a}_{ij}(\omega)$, 
one can use Eq.~(\ref{eq:16}) by substituting $\widetilde{\Sigma}_{il}^{r}(\omega)$ by the following 
advanced self-energy 
\begin{equation} 
\widetilde{\Sigma}_{il}^{a}(\omega)=-\,\frac{\rmi}{2} 
\langle X_i \rangle \langle X_l\rangle (\Gamma_{il}^L+\Gamma_{il}^R)\ .
\label{eq:17c}
\end{equation}
\end{subequations}

\bibliography{references}

\end{document}